\documentclass[pre,aps,twocolumn,superscriptaddress,showpacs,floatfix]{revtex4}
\usepackage{graphicx}
\usepackage{subfigure}
\usepackage{amsmath, amssymb}
\usepackage{epstopdf}
\newcommand{\bb}{\begin{equation}}
\newcommand{\en}{\end{equation}}

\begin{document}

\title{Diffusion and binding of finite-size particles in confined geometries}
\author{Mark L. Henle}
\affiliation{Department of Chemistry and Biochemistry, University of California, Los Angeles, CA 90025}
\author{Brian DiDonna}
\affiliation{Stellar Science, 6565 America's Parkway NE, Suite 725, Albuquerque, NM 87110}
\author{Christian D. Santangelo}
\affiliation{Department of Physics, University of Massachusetts, Amherst, MA  01003}
\author{Ajay Gopinathan}
\affiliation{School of Natural Sciences, University of California, Merced, CA 95344}

\date{\today}

\begin{abstract}
Describing the diffusion of particles through crowded, confined environments with which they can interact is of considerable biological and technological interest. Under conditions where the confinement dimensions become comparable to the particle dimensions, steric interactions between particles, as well as particle-wall interactions, will play a crucial role in determining transport properties.  To elucidate the effects of these interactions on particle transport, we consider the diffusion and binding of finite-size particles within a channel whose diameter is comparable to the size of the particles. Using a simple lattice model of this process, we calculate the steady-state current and density profiles of both bound and free particles in the channel. We show that the system can exhibit qualitatively different behavior depending on the ratio of the channel width to the particle size. We also perform simulations of this system, and find excellent agreement with our analytic results.
\end{abstract}


\maketitle

\section{Introduction}
\label{sec:intro}

Recent advances in technology and a burgeoning interest in biological systems have generated a great deal of interest in understanding particle transport in crowded environments.  There are many biological processes where the diffusion of particles through the crowded environment of the cell is important. Examples of these include: The transport of material through ion channels~\cite{ion, chou}, mitochondrial and bacterial porins~\cite{porin}, and nuclear pores~\cite{np}; and the diffusion of enzymes and macromolecules through microtubules and microtubule bundles~\cite{odde,deborah}.   In addition, cells themselves diffuse through confined environments, such as the movement of red blood cells and leukocytes through small blood vessels~\cite{rbc}.  Finally, the transport of organelles between cells has been shown to occur through narrow ``nanotubes'' that connect the cells~\cite{rustom}.  In technology, microfluidic devices and techniques~\cite{quake,stone} have immense potential for a variety of applications, including miniature biological assays for diagnostics and basic research~\cite{tay,makamba}.  Other applications where such considerations would be important include diffusion through carbon nanotubes~\cite{hummer} and microporous materials such as zeolites~\cite{chou, ruthven, kukla}, as well as the diffusion of colloidal particles through narrow channels~\cite{wei, lin}.

To date, much of the theoretical effort on particle diffusion in confined environments has focused on two extreme limits.  The first limit is the  ``single-file diffusion'' case, where the effects of confinement are so severe that steric interactions between the particles prevent them from diffusing past one another~\cite{chou, spitzer, levitt, liggett, derrida}.  In many systems, however, steric interactions are important, but the level of confinement is not so extreme, allowing the diffusion of a small number of particles past one another.  Attempts have been made to model such systems using both continuum~\cite{percus} and lattice-based approaches~\cite{kehr, kolomeisky, popkov}, but these studies address regimes close to the single-file limit using either perturbative, quasi-single-file models or two-file lattice models.  In the opposite limit, steric interactions between the particles are ignored, but other effects of the confining environment are taken into account.  For example, the friction between the particles and the confining walls can lead to hinderance of diffusion~\cite{odde}.  Also, the variation of the cross-sectional area of a channel has been shown to lead to a generalized one-dimensional diffusion equation known as the Fick-Jacobs equation~\cite{fick, jacobs, zwanzig, reguera, kalinay}.  Therefore, it is clear that understanding the general problem of confined diffusion, where the degree of confinement is lessened but steric interactions remain important, remains an important and under-explored endeavor.  

Another important effect in these confined systems that has not received extensive theoretical attention is the effect of specific and non-specific interactions between the particles and the confining environment.    For example, enzymes can bind to and chemically modify specific sites in tubulin when diffusing through microtubules~\cite{odde,deborah}. Also, microfluidic channel walls can be functionalized to allow for the binding of ligands in order to enable detection~\cite{tay,makamba}.  Furthermore, electrostatic and van der Waals forcese are non-specific particle-wall interactions that, when combined with the effects of surface roughness and charge inhomogeneity of the walls, can lead to localized, transient binding of the particles to the walls.  

In this paper, we develop a simple model to study the diffusion of finite size particles through narrow channels with functionalized walls to which the particles can reversibly bind. We consider the limit in which the channel width is a few times larger than the particle dimensions, so that the particles can diffuse past each other relatively easily.  On the other hand, the binding of particles on the channel walls can cause a bottleneck, effectively narrowing the dimensions of the channel for unbound particles.  In section \ref{sec:model}, we describe in detail the setup of our basic model and (in conjunction with Appendix A) the analytical procedure utilized to solve it. We also describe the simulations used to test our theoretical predictions.  In section \ref{sec:nobind} we describe the simplest case of diffusion in the absence of reversible binding, and make connections to both the  standard results for bulk diffusion and to the diffusion of particle through a channel of varying cross section~\cite{fick, jacobs, zwanzig, reguera, kalinay}. We then consider the effects of reversible binding on particle transport through the channel. We discuss two different cases, depending on the diameter of the channel relative to the particle size.  In section \ref{sec:nohole} we consider the case where the channel diameter is small enough that it can be completely blocked by bound particles.  We analyze the flux of particles through the channel and the densities of bound and unbound particles within the channel, and show that our simulation results are in good agreement with the analytical predictions. We also show that corrections to mean-field theory are necessary to account for the observed transport properties. In section \ref{sec:hole} we consider the case where the channel is too wide to be completely blocked by bound particles.  We show that the transport properties, which in this case are adequately described by the mean-field theory, are significantly modified relative to bulk diffusion and to the case considered in section~\ref{sec:nohole}.  We conclude with a discussion of our results and their implications for biological and technical applications.

\section{Microscopic Model}
\label{sec:model}

\begin{figure}
\includegraphics{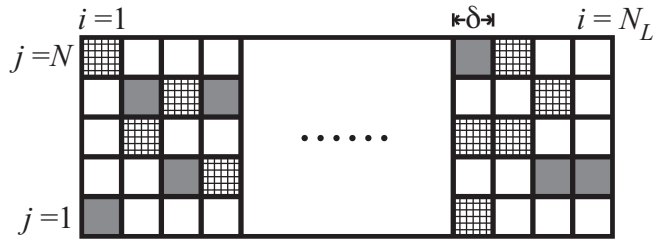}
\caption{\label{fig:Schematic} Schematic illustration of the interior of the channel for a typical particle distribution of bound particles (shaded sites), unbound particles (cross-hatched sites), and unoccupied (white) sites.} 
\end{figure}

Consider a channel of cross-sectional area $a_t$ in which particles of diameter $\delta$ can both diffuse and bind to the channel walls.  We partition the cross section into a number of bins labeled by the index $j$, as illustrated schematically in Fig.~\ref{fig:Schematic}.   The ratio $4a_t/(\pi \delta^2)$ determines the maximum number $N$ of particles that can fit within any given cross section of the channel; in our model, $N$ is the number of rows in the channel. We label sites along the axis of the channel with an integer $i=1,...,N_L$, where $N_L \equiv L/\delta$, $L$ being the length of the channel. The index $j$ ranges from $1,...,N$.  Due to the steric interactions between particles, each site $(i,j)$ can be occupied by at most one particle.  If a row in the model corresponds to a region adjacent to the walls of the channel, the particles can reversibly bind to the sites in this row.  By varying the ratio $w/\delta$, where $w \propto \sqrt{a_t}$ is the width of the channel, we can see that there are two distinct cases we need to consider.  When $w/\delta \sim 1$, every row $j$ is an ``exterior'' row that lies along the wall of the channel.  In this case, which we call the ``no-hole case,'' it is impossible for particles to diffuse through cross sections of the channel which contain the maximum number of bound particles.  When $w/\delta \gg 1$, however, there are a certain number $N_H$ of ``interior'' rows where particles cannot bind.  In this case, which we call the ``hole case,'' particles can always freely diffuse in the center of the channel, even in regions where the maximum number of particles are bound to the channel walls.  

For both of the scenarios described above, the system evolves forward in time as a Markov process.  That is, in a discrete time step $\Delta t$, each particle stochastically determines which (if any) of its possible moves it will attempt using a given state-independent probability for each move.  Due to steric interactions, however, the chosen move can occur only if the particle attempts to move to an unoccupied site.  If the site is occupied, the particle does not move in that time step.   There are several possible moves that each particle can attempt in a single time step.  In general, the diffusion of particles in a channel can be ``asymmetric''~\cite{chou, spitzer, levitt, liggett, derrida}, with different rates for hopping to the left and right.  Also, the diffusion constant can in principle vary with the distance from the channel walls~\cite{odde} (i.e. it can depend on the row index $j$).   In this paper, however, we assume that the diffusive landscape for the unbound particles in the channel is completely flat.  In other words, we consider the symmetric diffusion process exclusively.  Furthermore, this assumption implies a symmetry between the rows in the channel that requires the left and right hopping rates to be independent of the row index $j$.  Thus,  an unbound particle at any site $(i,j)$ can hop to the site $(i \pm 1, j)$ with probability $p_{\rm hop}$.  In addition,  an unbound particle at $(i,j)$ can hop to the site $(i, j')$ with probability $\tilde{p}_{\rm hop} (j,j')$.  In principle, the functional dependence of $\tilde{p}_{\rm hop} (j,j')$ on the rows $j,$ $j'$ must encapsulate the geometry of the channel. The determination of  these rates for arbitrary channel geometries could prove difficult, though the aforementioned symmetry between the rows requires $p_{\rm hop}(j, j') = p_{\rm hop}(j',j)$. Fortunately, we shall see that a detailed knowledge of these rates will not be necessary to find the quantities of interest in this paper.  Finally, if the index $j$ labels an exterior row, an unbound particle at $(i,j)$ can bind to that site with probability $p_{\rm on}$, while a bound particle can unbind with probability $p_{\rm off}$ .  

To determine the particle profiles, we need to specify the boundary conditions at either end of the channel. Throughout this paper, we place the left end of the channel in contact with a bath of particles in the bulk, which sets the number of particles at $i=1$.  At the right end of the channel, we place an absorbing boundary (i.e. an infinitely dilute bath), which forces the particle profiles to vanish at $i=N_L$.  

Analytically, the time evolution of this system can be described by the master equation:
\bb
\label{eq:masterEq}
\left| \psi \left(t+ \Delta t\right)\right> - \left| \psi (t)\right> \equiv  \left| \Delta \psi (t)\right> = K \left| \psi (t)\right>,
\en
where $K$ is the evolution operator and $\left| \psi (t) \right>$ is the ``state vector'' of the system at time $t$:
\bb
\label{eq:psiDef}
\left|  \psi (t)\right> \equiv \sum_{\{u_{i,j} \}=0,1} \sum_{\{b_{i,j} \}=0,1} P\left[ s ;t \right] \left|s\right>.
\en
Here,  $u_{i,j}$ ($b_{i,j}$) is the number of unbound (bound) particles at site $(i,j)$,  the ``eigenstate'' $\left|s\right> \equiv \left| \{u_{i,j}, b_{i,j}\}\right>$ enumerates one specific configuration for all of the sites in the channel, and $P[s ;t ]$ is the probability of finding the system in the eigenstate  $\left|s\right>$ at time $t$.  The derivation of the evolution operator $K$ is given in Appendix \ref{sec:masterEq}.   We note that although the sum in Eq.~(\ref{eq:psiDef}) includes physically unallowable states (specifically, states with multiple particles occupying the same site), we can always force the probability of such states to vanish at all times (see Appendix~\ref{sec:masterEq}).  

Using the evolution operator $K$, we can compute the time evolution of the expectation value (i.e. the thermal average) of any physical observable.  In this paper, we will focus excusively on the variation of physical observables \emph{along} the channel axis, rather than their variation \emph{within} a given cross section of the channel.  In particular, we are interested in the total number of bound and unbound particles  in the channel as a function of the (dimensionless) distance $i$ along the channel axis, as well as the current, i.e. the total number of particles per unit time passing through a given cross section of the channel.  These quantities can be related to the expectation values of two operators, $C_b (i, j)$ and $C_u (i,j)$: $C_b (i, j)$ gives $1$ if the site $(i, j)$ is occupied by a bound particle, and $0$ otherwise, while $C_u (i, j)$ gives $1$ if the site $(i, j)$ is occupied by a unbound particle, and $0$ otherwise. The evolution equations for these operators are derived in Appendix~\ref{sec:masterEq}.  

In order to test the validity of our analytic results, we have performed simulations of the lattice model described above. All simulations were done on a rectangular lattice like the one illustrated in Fig.~\ref{fig:Schematic}, with $N=5$ and $N_L =100$.  For simplicity, we set all of the lateral and longitudinal hopping probabilities to be equal, $\tilde{p}_{\rm hop} (j,j') = \tilde{p}_{\rm hop}$.
For every time step in the simulation, each particle in the channel is visited once, in a random order. During each particle visit, it is first determined what (if any) move that particle will attempt, using the rules and probabilities defined above.  If that move is allowed -- that is, if it does not lead to any multiply occupied sites in the lattice -- then it is performed; if that move is not allowed, then the attempt fails.    This process is then repeated for the next (randomly chosen) particle, until all of the particles have been visited once during the time step. 

The boundary conditions in the simulation are set as follows:  First, particles that leave either end of the channel do not return.  This alone sets the absorbing boundary condition at the right end of the channel.  To set the boundary condition at the left end of the channel, we need an influx of particles into the channel at that end. To provide this influx, we stochastically attempt -- with probability $p_{\rm enter}$ -- to insert a single additional particle into the left end of the channel at the end of each time step. If it is determined that an attempt should be made, then one of the $N$ sites in the column $i=1$ is chosen at random as the particle entry point. If that site is empty then the new particle is added there; if the site is occupied then the attempt fails.  The value of the probability  $p_{\rm enter}$ sets the number of particles in the column $i=1$, which must be measured in order to compare the simulation results to the theoretical solutions.

\section{Diffusion without binding}
\label{sec:nobind}

Before considering the full problem of diffusion and binding of finite-size particles inside a channel, we first consider the limit $p_{\rm on}, p_{\rm off} \rightarrow 0$, in which the diffusing particles cannot reversibly bind to the channel walls. It is possible, however, to have an initial, stationary distribution of irreversibly bound particles in this limit.  

We first consider an initial condition with no bound particles.  In this case, the problem reduces to the diffusion of finite-size particles with excluded volume interactions through a channel with a uniform cross section. Here, the only quantity of interest is the number of particles along the channel, $N_u (i, t) \equiv \sum_j \left<C_u (i,j)\right>$.  Using the results of Appendix~\ref{sec:masterEq}, it is straightforward to show that,
in the continuum limit ($\Delta t, \delta \rightarrow 0$) Eq.~(\ref{eq:CuEvolve}) yields the standard diffusion equation for phantom (i.e. point-like) particles with: 
\bb
\label{eq:diffusionNoBind}
\partial_t \lambda_u (x,t) = D \lambda_u''(x,t),
\en
where $x \equiv i \delta$ is the continuous position along the channel and  $D = \lim_{\Delta t, \delta \rightarrow 0}  p_{\rm hop} \delta^2/\Delta t$ is the diffusion constant.  Here, $\lambda_u (x, t) \equiv \lim_{\delta \rightarrow 0} N_u (i, t)/ \delta$ is the number of particles per unit length at position $x$; that is, $\lambda_u (x, t) \, dx$ is the number of particles between $x$ and $x+dx$.

The fact that excluded volume interactions do not alter the simple diffusive behavior Eq.~(\ref{eq:diffusionNoBind}) of the particle profile is due to the assumed symmetry of the particle diffusion constant along the channel axis.  Indeed, it is well known that excluded volume interactions do not affect the bulk diffusion equation when the diffusion constant is independent of position, even in the single-file limit (i.e. the symmetric exclusion process) \cite{derrida2}.  In the case of an asymmetric exclusion process, where the hopping rate from $i$ to $i+1$ is different from the hopping rate from $i$ to $i-1$, excluded volume interactions do indeed affect the bulk diffusion equation \cite{schultz}.  Furthermore, excluded volume interactions do play a role in the behavior of individual particles in the channel (i.e.  tracer diffusion), even for symmetric diffusion processes \cite{derrida2}.  Finally, we note that the terms in Eq.~(\ref{eq:CuEvolve}) for hopping within a given column $i$ -- i.e. the terms $\propto \tilde{p}_{\rm hop} (j,j')$ -- cancel exactly in Eq.~(\ref{eq:diffusionNoBind}).  This occurs for \emph{arbitrary} values of the lateral hopping rates $ \tilde{p}_{\rm hop} (j,j')$, as long as these rates are symmetric, $ \tilde{p}_{\rm hop} (j,j') =  \tilde{p}_{\rm hop} (j',j)$.  Thus, the lateral diffusion of the particles \emph{within} the channel has no effect on the particle profile \emph{along} the channel axis.  

We can also use the limit  $p_{\rm on}, p_{\rm off} \rightarrow 0$ to study the diffusion of particles through a channel with a cross section that varies on length scales much longer than the particle size. To do so, we choose initial conditions such that $\left<C_b (i, j)\right> = n_b (i,j)$ is a fixed function that represents the varying cross section of the channel.  Then Eq.~(\ref{eq:CuEvolve}) becomes
\begin{align}
\label{eq:FJdiscreteNH}
\Delta N_u (i,t) = p_{\rm hop} \sum_\pm \sum_j & \left[ \left< n_u(i\pm1,j) - n_u(i,j) \right>\right] \\
& \times [1-n_b(i\pm1,j)] [1-n_b(i,j)].\nonumber
\end{align}
 
If we assume that the cross section varies on length scales much longer than the channel radius, we can approximate the distribution of particles  within a given cross section of the channel by a uniform distribution.  This is known as the ``local equilibrium approximation''~\cite{zwanzig}. In this limit,
\bb
\label{eq:localEq}
\left< n_u(i,j) \right> = \frac{N_u(i)}{[N - N_b(i)]},
\en
where $N_b(i) \equiv \sum_j n_b (i,j)$.
Using Eq.~(\ref{eq:localEq}), Eq.~(\ref{eq:FJdiscreteNH}) becomes
\bb
\label{eq:FJdiscreteNH2}
\Delta N_u (i,t) = p_{\rm hop} \sum_\pm \left[\frac{N_u(i \pm 1)}{A (i\pm1)} -\frac{N_u(i)}{A(i)}\right] Q(i\pm1),
\en
where the ``cross-sectional area'' $A(i) \equiv N-N_b (i)$ is the number of particles that can occupy row $i$, and the ``permeability'' $Q(i\pm 1) \equiv \sum_j [1-n_b (i\pm 1,j)][1-n_b(i,j)]$.  Now, the assumption that the cross section of the channel varies on length scales much longer than the particle size $\delta$ implies that the function $n_b (i, j)$ is a slowly varying function of the row index $i$.  Therefore, in the continuum limit $n_b(i \pm 1,j) = n_b(i,j) + \mathcal{O}(\delta)$, and the permeability is given by
\begin{align}
\label{eq:Qapprox}
\notag
Q(i \pm 1) &= \sum_j\left[1-n_b(i,j)\right]^2 + \mathcal{O}(\delta)\\
&=  \sum_j\left[1-n_b(i,j)\right] + \mathcal{O}(\delta) = A (i)+ \mathcal{O}(\delta).
\end{align}
The second equality follows from the fact that $1-n_b(i,j)=0,1$ always.  Then in the continuum limit Eq.~(\ref{eq:FJdiscreteNH2}) becomes
\bb
\label{eq:diffusionFJ}
\frac{\partial \lambda_u(x,t)}{\partial t} = - J'[x,t],
\en
where the prime indicates a derivative with respect to the spatial variable $x$, and the current
\bb
\label{eq:FJcurrent}
J (x,t) = -D \left\{A(x) \frac{\partial}{\partial x} \left[ \frac{\lambda_u(x,t)}{A(x)} \right] \right\}.
\en
Here, $A(x) \equiv \lim_{\delta \rightarrow 0} A(i)/\delta$ is  the  continuum cross-sectional area; that is, $A(x) dx$ is the maximum number of particles that can simultaneously occupy the region in the channel between $x$ and $x+dx$.  We note that the $\mathcal{O}(\delta)$ terms of the permeability Eq.~(\ref{eq:Qapprox}) vanish in the continuum limit $\delta \rightarrow 0$.  

Eqs.~(\ref{eq:diffusionFJ}) and~(\ref{eq:FJcurrent}) are known as the Fick-Jacobs equation, and have already been derived from the continuum diffusion equation for a \emph{cylindrically symmetric} channel~\cite{fick, jacobs, zwanzig, reguera, kalinay}.  Our result generalizes the validity of the Fick-Jacobs equation to any channel with cross-sectional area $A(x)$.  In particular, a channel with changing area due to a change in the \textit{shape} of the cross section will also exhibit Fick-Jacobs behavior if the shape change occurs slowly enough.

\section{No-Hole Case}
\label{sec:nohole}

We now turn to the case in which particles can reversibly bind to the walls of the channel. In this section, we consider the ``no-hole'' case, in which the diffusion of unbound particles through a region of the channel can be completely blocked by bound particles in that region.  In the language of the lattice model, particles can bind to every site $(i,j)$ of the channel. The state of this system can be described by two functions, $N_\alpha(i) = \sum_j \left<C_\alpha(i,j)\right>$, which give the expected number of bound ($\alpha=b$) and unbound ($\alpha = u$) particles at position $i$ along the channel axis.  If we take the continuum limit, the evolution equations~(\ref{eq:CbEvolve}) and~(\ref{eq:CuEvolve}) become, respectively, 
\bb
\label{eq:BoundEvolve}
\frac{\partial}{\partial t} \lambda_b(x,t) = k_{\rm on} \lambda_u(x,t) - k_{\rm off} \lambda_b(x,t), 
\en
\bb
\label{eq:UnboundEvolveNH}
\frac{\partial}{\partial t} \lambda_u(x,t)= - k_{\rm on} \lambda_u(x,t) + k_{\rm off} \lambda_b (x,t)-J'(x,t),
\en
where $k_{\rm on, off}=p_{\rm on, off}/\Delta t$ and $\lambda_\alpha (x, t) = \lim_{\delta \rightarrow 0} N_\alpha (i, t)/\delta$.  The discrete longitudinal current $\bar{J} (i, t)$ is given by
\begin{align}
\label{eq:JdiscreteNH}
\bar{J} &(i, t) \equiv\frac{p_{\rm hop}}{\Delta t} \Big[ N_u(i+1)- N_u(i)\Big.
\\
\nonumber
&\Big. +\sum_j  \big(\left<C_u(i,j) C_b(i+1,j)\right>-\left<C_u(i+1,j) C_b(i,j)\right>\big)\Big].
\end{align}
The continuous current in Eq.~(\ref{eq:UnboundEvolveNH}) is related to the discrete current by $J (x,t) =  \lim_{\delta \rightarrow 0} \bar{J} (i,t)$.  Like the simple diffusion case, the terms for hopping within a given column $i$ cancel for arbitrary values of the hopping rates $\tilde{p}_{\rm hop} (j, j')$ as long as $\tilde{p}_{\rm hop} (j, j')=\tilde{p}_{\rm hop} (j', j)$. The first two terms of Eq.~(\ref{eq:JdiscreteNH}) 
give the lattice version of the current for phantom particles (i.e. Fick's Law), while the final two terms give the correction due to the fact that an unbound particle at $(i,j)$ cannot hop to a site $(i \pm 1, j)$ if it is occupied by a bound particle.

As mentioned in Section~\ref{sec:model}, our model assumes that the diffusive landscape for the unbound particles in the channel is completely flat.  In the no-hole case, this implies that any two rows of the channel are interchangeable.  As a result, the steady state expectation value of any operator that acts on a single row $j$ will be independent of $j$, as long as the boundary conditions are independent of $j$. In more detail, consider the probability $P[s;t]$ of a particular configuration $s$ of bound, occupied, and unoccupied particles in the channel arising at time $t$. After interchanging two rows $j$ and $j'$, we arrive at a configuration $s'$ of particles that can arise with probability $P[s';t]$ at time $t$. Symmetry between the rows implies that $P[s;t] = P[s';t]$. When this symmetry is present,
\begin{align}
\label{eq:NoHoleSymm}
\left< C_\alpha (i,j)\right> &= \left< C_\alpha (i) \right> ,\\
\notag
\left< C_\alpha (i,j) C_{\alpha'} (i',j) \right> &= \left< C_\alpha (i) C_{\alpha'}(i') \right>, \quad \alpha, \alpha'=u, b.
\end{align}
That is, averages involving operators of a single row $j$ must be independent of $j$, since those averages involve summing over the probability distributions $P[s;t]$ described above. Thus, we obtain the same expressions for the expectation value of any operator $\left< \mathcal{O}(j) \right>$ after interchanging $j$ with $j'$, implying Eq.~(\ref{eq:NoHoleSymm}) directly.

In order to solve Eq.~(\ref{eq:UnboundEvolveNH}), we must postulate a form for the two-point correlation functions appearing in Eq.~(\ref{eq:JdiscreteNH}).  The simplest form for these correlation functions is given by the mean-field approximation, in which the correlations between the two operators are neglected:
\bb
\label{eq:MFapprox}
\left< C_u (i,j) C_b(i',j) \right> = \left<C_u (i) \right> \left< C_b(i') \right>.
\en
Then the mean-field current, in the continuum limit, becomes
\begin{align}
\nonumber
\label{eq:Jmf}
J_{\rm mf} (x,t) =& \frac{D}{\Lambda}\left[\lambda_u' (x,t) \lambda_b (x,t) - \lambda_u (x,t) \lambda_b' (x,t)\right]\\
&-D \lambda_u'(x,t),
\end{align}
where $\Lambda \equiv \lim_{\delta \rightarrow 0} N/\delta$ is the maximum number of particles (both bound and unbound) that can fit in the channel, per unit length,  

At steady state, $\partial_t \lambda_u (x,t) = \partial_t \lambda_b (x,t) =0$ and the solution to Eq.~(\ref{eq:BoundEvolve}) is
\bb
\label{eq:UnboundSS}
\lambda_u (x) = K_{\rm d} \lambda_b (x),
\en
where the disassociation constant $K_{\rm d} \equiv k_{\rm off}/k_{\rm on}$.  Note that this solution is independent of the mean-field approximation, Eq.~(\ref{eq:MFapprox}).  Using Eqs.~(\ref{eq:Jmf}),~(\ref{eq:UnboundSS}), and the boundary conditions $\lambda_u (L) = \lambda_b (L) =0$, $\lambda_b (0) \equiv \lambda_b^{(0)}$, the steady state solution to Eq.~(\ref{eq:UnboundEvolveNH}) is a simple linear profile:
\bb
\label{eq:BoundMFTnoHole}
\lambda_b^{\rm mf} (x) = \lambda_b^{(0)}\left[1- \frac{x}{L}\right],
\en
and the steady state current is
\bb
\label{eq:JmfSS}
J_{\rm mf} (x) = - D \lambda'_u (x) = \frac{\lambda_b^{(0)} K_{\rm d} D}{L} \equiv J_0^{\rm mf}.
\en

\begin{figure}
\includegraphics{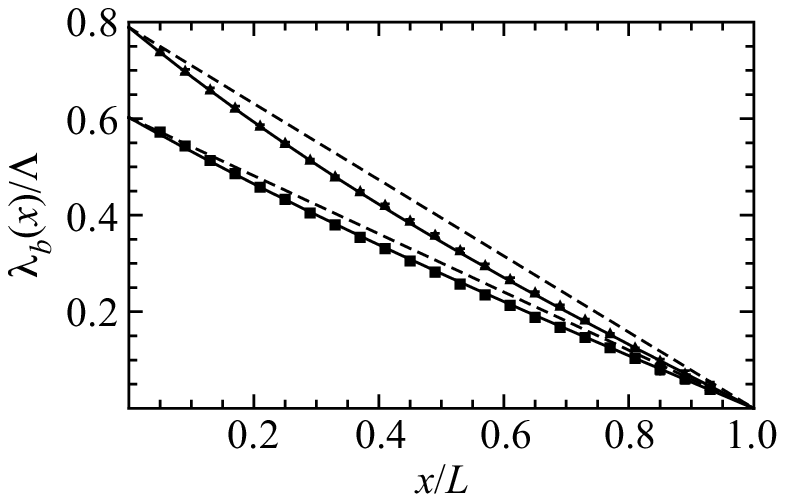}\vspace{10pt}
\includegraphics{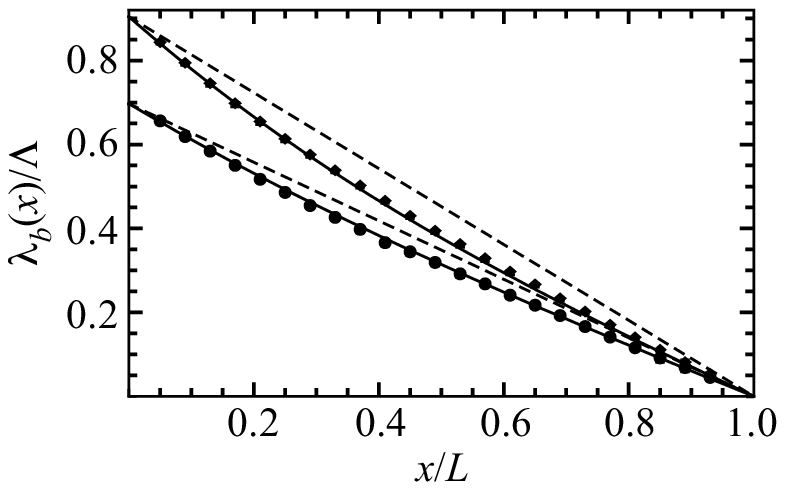}
\caption{\label{fig:ProfileNH} Dimensionless bound particle density $\lambda_b (x)/\Lambda$ for the no-hole case obtained from the simulations (points), as compared to the theoretical predictions both at the mean-field level (dashed lines) and including the correlations (solid lines).  For all data shown, $N=5, \, p_{\rm on} = 0.01, \,p_{\rm off} = 0.001$, and $p_{\rm hop} = 1/150$; for the simulations, we also set $\tilde{p}_{\rm hop} (j, j') = p_{\rm hop}$ for all $j, j'$. The value of $p_{\rm enter}$ for each simulation sets the left-end boundary condition $\lambda_b^{(0)}$ for the theoretical curves.  The values of $p_{\rm enter}$ and $\lambda_b^{(0)}$, respectively, are: $0.006,$ $0.60$ (\emph{top}, squares); $0.02,$ $0.79$ (\emph{top}, triangles); $0.01,$ $0.70$ (\emph{bottom}, circles); and $0.5,$ $0.90$ (\emph{bottom}, diamonds). For the theoretical predictions that include the effects of correlations, all of the curves use the same fitting parameter, $\epsilon = 0.27$.} 
\end{figure}

\begin{figure}
\includegraphics{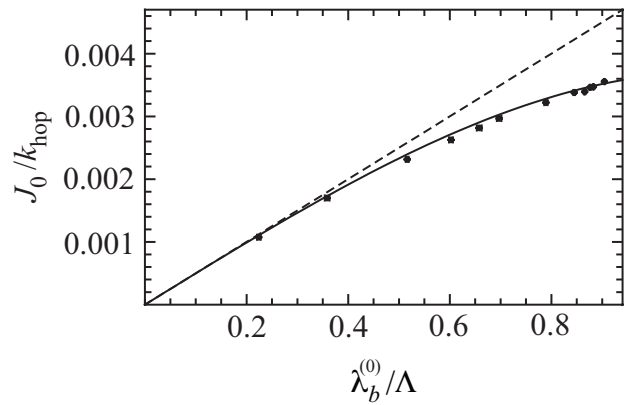}
\caption{\label{fig:CurrentNH} Dimensionless steady-state current $J_0/k_{\rm hop}$ obtained from the simulations (points) for various values of the left-end boundary condition $\lambda_b^{(0)}/\Lambda$, as compared to the theoretical predictions both at the mean-field level (dashed lines) and including the correlations (solid lines).  All parameter values are identical to those used in Fig~\ref{fig:ProfileNH}.}
\end{figure}

Figures~\ref{fig:ProfileNH} and~\ref{fig:CurrentNH} show the bound particle steady-state profile $\lambda_b (x)$ and the the steady-state current $J_0$, respectively.  In both figures, we can see distinct deviations of the simulations (points) from the predicted mean-field predictions (dashed lines). These deviations arise from the mean-field treatment of correlation functions $\left<C_u (i,j) C_b (i', j)\right>$.  In order to understand the physical processes that cause these deviations, let us first consider a quenched distribution of bound particles where the unbound particle density has reached a steady state.  In this case, there will be an absence of correlations between the bound and unbound particle distributions, because the bound particle distribution is invariant in time. However, if we now consider a single binding or unbinding event, it is clear that there will be a transient change in the surrounding unbound particle density as it  relaxes toward a new steady state distribution. Since particles bind and unbind on finite timescales, every such event will lead to a transient deviation in the correlation functions from their mean field value. This deviation will be particularly significant if a binding (unbinding) event blocks (unblocks) the entire cross section of the channel.

To quantify the deviations from mean-field theory, we need to construct a dimensionless quantity that involves the two-point correlation functions  appearing in Eq.~(\ref{eq:JdiscreteNH}). Although the operators $C_\alpha (i,j)$ are dimensionless in their discrete form, their expectation values in the continuum limit $\lambda _\alpha (x)$ have units of $\left(length\right)^{-1}$.  Therefore, we characterize the deviations of the expectation value $\left<C_u (i \pm 1,j) C_b (i, j)\right>$ from its mean-field value with the dimensionless quantity
\bb
\chi_\pm (i) \equiv \frac{\left<C_u (i \pm 1,j) C_b (i, j)\right> - \left<C_u (i \pm 1,j) \right>\left<C_b (i, j)\right>}{ \left<C_u (i \pm 1,j) \right>\left<C_b (i, j)\right>}.
\en
Here, we have indicated the independence of the function $\chi_\pm$ from the row index $j$, which, as argued above, is known by symmetry.  Physically we anticipate that $\chi_\pm$ will be a function of the dimensionless density of the bound and unbound particles. Assuming the deviations are small, we may expand $\chi_\pm(i)$ in powers of $N_b/N$ and $N_u/N$ and find, to lowest order,
\bb
\label{eq:MFTdev}
\chi_\pm (i) = \epsilon_1 \frac{N_b (i)}{N} + \epsilon_2 \frac{N_u (i \pm 1)}{N}.
\en
Here, $\epsilon_1$ and $\epsilon_2$ are parameters that encapsulate the degree of deviation from mean-field behavior. Note that these parameters are not universal: in principle, they depend on the various hopping rates. 

Given Eq.~(\ref{eq:MFTdev}), the current can be written as $J (x, t) = J_{\rm mf} (x, t) +J_{\rm corr} (x, t)$, where $J_{\rm mf} (x, t)$ is given by Eq.~(\ref{eq:Jmf}) and 
\begin{align}
\label{eq:Jcorr}
\notag
J_{\rm corr} &(x, t) = -\frac{D \epsilon_1}{\Lambda^2}\left[\lambda_u(x, t) \partial_x \lambda_b^2 (x, t)- \lambda_b^2 (x, t) \lambda_u' (x,t)\right]\\
& + \frac{D \epsilon_2}{\Lambda^2}\left[\lambda_b (x, t)\partial_x \lambda_u^2 (x, t)- \lambda_u^2 (x, t) \lambda_b' (x,t)\right].
\end{align}

The steady-state relation given by Eq.~(\ref{eq:UnboundSS}) still holds, since (as noted above) it is independent of the mean-field approximation.  However, the steady-state bound particle profile and current are now given by, respectively,
\bb
\label{eq:BoundCorrNoHole}
\lambda_b (x) \approx \lambda_b^{\rm mf} (x) \left[1-\epsilon \left(\frac{\lambda_b^{(0)}}{\Lambda}\right)^2 \frac{x}{L} \left(2-\frac{x}{L}\right)\right],
\en
and
\bb
\label{eq:JbeyondMF}
J(x, t) = J_0 = \frac{D K_{\rm d} \lambda_b^{(0)}}{L}\left[1-\epsilon \left(\frac{\lambda_b^{(0)}}{\Lambda}\right)^2\right],
\en
where $\epsilon \equiv (-\epsilon_1 + K_{\rm d} \epsilon_2)/3$. Here, we have only retained the terms linear in $\epsilon$ in Eq.~(\ref{eq:BoundCorrNoHole}), in order to be consistent with the expansion of the correlation functions Eq.~(\ref{eq:MFTdev}).  Thus, we can see that our proposed form of the deviations from mean-field behavior has only one fitting parameter, $\epsilon$. 

Figures~\ref{fig:ProfileNH} and~\ref{fig:CurrentNH} show the resultant fits (solid lines) of Eqs.~(\ref{eq:BoundCorrNoHole}) and~(\ref{eq:JbeyondMF}), respectively, to the simulation data .  In order to determine the value of the fitting parameter $\epsilon$, we fit Eq.~(\ref{eq:JbeyondMF}) to the current data shown in Fig.~\ref{fig:CurrentNH}, and then use this value for all of the particle profiles shown in Fig.~\ref{fig:ProfileNH}. We can see that this gives excellent fits for all of the profiles and for the current measured by the simulations.  Thus, our postulate for the form of the deviations from mean-field theory captures the deviations seen in the simulations using only a single parameter fit.

\section{Hole Case}
\label{sec:hole}

We now turn to the case in which the ratio of the channel diameter to the particle size is large enough that  particles can always diffuse through a ``hole''  in the center of the channel, even when all of the binding sites in a given cross section of the channel are occupied by other bound particles.  In our lattice model, this corresponds to two distinct types of rows:  exterior rows, in which particles can diffuse and reversibly bind, and interior rows, in which the particles can only diffuse.  Thus, the state of the system can be described by three operators:  $N_\alpha(i) = \sum_j \left< C_\alpha(i,j)\right> \theta_j$ for $\alpha = u, b$ gives the expected number of bound ($\alpha=b$) or unbound ($\alpha = u$) particles in the exterior rows of column $i$; $N_h (i) = \sum_j \left<C_u (i,j)\right> (1-\theta_j)$ gives the expected number of unbound particles in the interior rows of column $i$.  Here, $\theta_j=1$ for the exterior rows, and $0$ for the interior rows.    

For both the case of diffusion without reversible binding and the no-hole case considered above, the lateral hopping terms in the evolution equations of interest cancelled one another exactly.  This cancellation occurred for arbitrary values of the lateral hopping probabilities $\tilde{p}_{\rm hop} (j, j')$, subject only to the symmetry requirement $\tilde{p}_{\rm hop} (j, j') = \tilde{p}_{\rm hop} (j', j)$.  As we shall see below, however, the distinction between the interior and exterior rows in the hole case causes some of the lateral hopping terms -- specifically, those terms corresponding to the diffusion of particles from the interior rows to the exterior rows (and vice-versa) -- to remain in the relevant evolution equations.    Consequently, we need to make a simplifying assumption about these hopping probabilities.  Since the primary focus of this paper is the effects of steric interactions on the diffusion of particles in confined geometries, we need not consider cases where the number $N_{\rm H}$ of interior rows is large; the particle diffusion in such systems can be adequately described by the standard diffusion equation for phantom particles. Geometrically speaking, when $N_{\rm H}$ is small, all of the exterior rows will be approximately equidistant from any given interior row.  Therefore, we will assume that the probability of hopping from \emph{any} exterior row to \emph{any} interior row (and vice-versa) is given by $\tilde{p}_{\rm hop}=const$.   This assumption does not apply to the hopping probabilities from one interior row to another interior row, or from one exterior row to another exterior row:  these rates remain arbitrary, except for the usual the symmetry constraint  $\tilde{p}_{\rm hop} (j, j') = \tilde{p}_{\rm hop} (j', j)$.  Thus, we assume that the lateral hopping rates are of the form
\bb
\label{eq:PhopTildeHole}
\tilde{p}_{\rm hop} (j, j') = \tilde{p}_{\rm hop} \left[1- \Theta_{j, j'}\right] +\phi (j, j') \Theta_{j, j'}
\en
where $\phi (j, j') = \phi (j', j)$ and $\Theta_{j, j'}=0$ if $\theta_j \neq \theta_{j'}$ -- that is, if one row is an exterior row and the other is an interior row -- and $1$ if $\theta_j = \theta_{j'}$.  

If we take the continuum limit, the evolution equation for the bound particle profile reduces to the same equation obtained in the no-hole case, Eq.~(\ref{eq:BoundEvolve}).   We can use the evolution equation for $\left<C_u (i,j)\right>$, Eq.~(\ref{eq:CuEvolve}), to obtain the evolution equations for both of the profiles $N_u (i, t)$ and $N_h (i, t)$.  Specifically, if we use the assumption Eq.~(\ref{eq:PhopTildeHole}) for the lateral hopping rates, it is straightforward to show the terms $\propto \Theta_{j, j'}$ cancel one another exactly in both evolution equations, so that 
\begin{align}
\label{eq:HoleSymm}
\left<C_b (i,j)\right> &= \frac{N_b (i)}{N-N_H} \theta_j,\\
\left<C_u (i,j)\right> &= \frac{N_u (i)}{N-N_H} \theta_j+\frac{N_h (i)}{N_H} \left(1-\theta_j\right).
\end{align}
In contrast to our discussion of the no-hole case, we cannot exchange an interior row with an exterior row, although we can interchange interior (and exterior) rows amongst themselves. This latter symmetry ensures that the expectation values take on only one value on all interior rows and another on all exterior rows. Using Eqs.~(\ref{eq:PhopTildeHole}) and~(\ref{eq:HoleSymm}), it is straightforward to show that Eq.~(\ref{eq:CuEvolve}) gives the mean-field evolution equations
\begin{widetext}
 \bb
 \label{eq:UnboundEvolveHole}
\frac{\partial}{\partial t} \lambda_u(x,t) = -k_{\rm on} \lambda_u (x,t) + k_{\rm off} \lambda_b (x,t)
- \tilde{D} \Lambda \left[\lambda_u (x,t) \Lambda_H - \lambda_h (x,t) \left(\Lambda-\Lambda_H-\lambda_b (x,t) \right) \right]- J'_u (x,t),
\en
\bb
\label{eq:HoleEvolve}
\frac{\partial}{\partial t} \lambda_h (x,t) =  \tilde{D} \Lambda \left[\lambda_u (x,t) \Lambda_H - \lambda_h (x,t) \left(\Lambda-\Lambda_H-\lambda_b (x,t) \right) \right]- J'_h (x,t),
\en
where $\lambda_\alpha (x, t) \equiv \lim_{\delta \rightarrow 0} N_\alpha (i, t)/\delta$ for $\alpha = u, h, b$, $\Lambda_H =  \lim_{\delta \rightarrow 0} N_H/\delta$, $\tilde{D} \equiv \lim_{\Delta t, \delta \rightarrow 0} \tilde{k}_{\rm hop} \delta^2/(N \Delta t)$, and the currents
\bb
J_u (x, t) \equiv -D\left[\lambda_u'(x,t)+\frac{1}{\Lambda-\Lambda_H}\left(\lambda_u(x,t) \lambda_b'(x,t) - \lambda_b(x,t) \lambda_u'(x,t)\right) \right], \qquad J_h (x, t) \equiv -D \lambda_h' (x, t).
\en
\end{widetext}

At steady state, we can see that Eq.~(\ref{eq:UnboundSS}) still holds, and that the total longitudinal current $J_{\rm tot} (x) \equiv J_u (x) + J_h (x)$ is constant:
\bb
J_{\rm tot} (x) = - D \left[ \lambda'_u (x) +\lambda'_h (x)\right] = J_0.
\en
This, along with the boundary condition $\lambda_\alpha (L) = 0$  (for $\alpha =u,b,h$), implies that
\bb
\label{eq:NhSS}
\lambda_h (x) = \frac{J_0}{D}\left(L-x\right) - K_{\rm d} \lambda_b (x).
\en

To solve for the steady-state current $J_0$, we need an additional boundary condition relating the particle profiles in the exterior and interior rows. Outside of the channel, there is no net flux of particles in the direction perpendicular to the axis of the channel.  Since the current must be continuous across the boundaries of the channel, the lateral current at the channel ends must vanish.  Specifically, the rate of particles hopping from the interior to the exterior rows must equal the rate of particles hopping from the exterior to the interior rows at the channel ends; that is, the term in brackets in Eq.~(\ref{eq:HoleEvolve}) must vanish there.  This is trivially satisfied at the right end at steady state, since all of the particle profiles vanish there.  At the left end, we must have
\bb
\label{eq:noholeBC}
K_{\rm d} \lambda_b (0) \Lambda_H = \lambda_h (0) \Big(\Lambda-\Lambda_H - \lambda_b (0)\Big).
\en
Such an equation is not necessary in the no-hole case, since in that case symmetry dictates that the lateral current vanishes everywhere.  Using Eq.~(\ref{eq:noholeBC}), the steady state current is given by 
\bb
\label{eq:CurrentHole}
J_0 = \frac{K_{\rm d} D}{L} \left[\frac{\Lambda-\lambda_b (0) }{\Lambda-\Lambda_H-\lambda_b (0) }\right] \lambda_b (0) 
\en

The remaining ODE for the steady state profiles can be obtained by combining Eqs.~(\ref{eq:UnboundSS}),~(\ref{eq:HoleEvolve}), and~(\ref{eq:NhSS}):
\begin{align}
\label{eq:BoundProfileHole}
& K_{\rm d} \lambda''_b (x) = \tilde{D} \Lambda \Bigg[ K_{\rm d}  \lambda_b (x) \Lambda_H \Bigg.\\
\nonumber
& \Bigg.- \Big(\frac{J_0}{D} (L-x) - K_{\rm d} \lambda_b (x)\Big)\Big(\Lambda-\Lambda_H -\lambda_b (x)\Big)\Bigg].
\end{align}
This nonlinear ODE must be solved numerically using the boundary conditions $\lambda_b (0) = \lambda_b^{(0)}$ and $\lambda_b (L) = 0$.   

\begin{figure}
\includegraphics{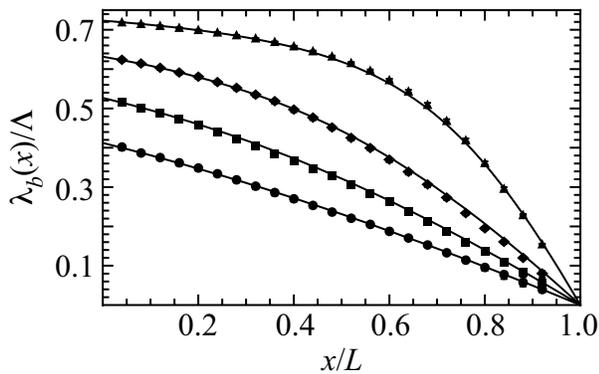}
\caption{\label{fig:ProfileH} Dimensionless bound particle density $\lambda_b (x)/\Lambda$ for the hole case obtained from the simulations (points), as compared to the mean-field theoretical prediction (solid lines).  For all data shown, $N_H=1$; the remaining parameter values are identical to those used in Fig.~\ref{fig:ProfileNH}.  The values of $p_{\rm enter}$, and the resultant values of $\lambda_b^{(0)}$, are, respectively:  $0.004,$ $0.41$ (circles); $0.008,$ $0.53$ (squares); $0.02,$ $0.63$ (diamonds); and $0.5,$ $0.72$ (triangles). } 
\end{figure}

\begin{figure}[t]
\includegraphics{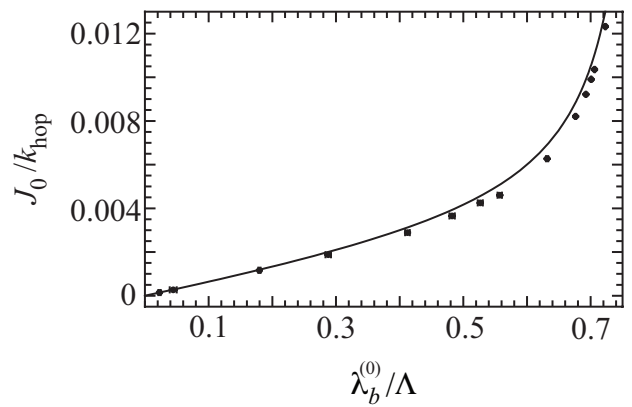}
\caption{\label{fig:CurrentH}  Dimensionless steady-state current $J_0/k_{\rm hop}$ obtained from the simulations (points) for various values of the left-end boundary condition $\lambda_b^{(0)}/\Lambda$, as compared to the mean-field theoretical predictions Eq.~(\ref{eq:CurrentHole}) [solid lines].  All parameter values are identical to those used in Fig~\ref{fig:ProfileH}.}
\end{figure}

Figures~\ref{fig:ProfileH} and~\ref{fig:CurrentH} show the simulation results (points)  for several steady-state bound particle profiles $\lambda_b (x)$ and the the steady-state current $J_0$, respectively, for the hole case.  We can see that the mean-field predictions (solid lines) for the particle profiles and the current show excellent agreement with the simulation results with no fitting parameters.  Thus, in contrast to the no-hole case, the effects of the correlations that are ignored in the mean-field approximation are negligible in the hole case.  Specifically, we can see from original evolution equation, Eq.~(\ref{eq:CuEvolve}), that the relevant two-point correlation functions are of the form $\left< C_u (i, j) C_b (i',j')\right>$.  In the no-hole case, the correlation of these two operators can be significant, because the hopping of unbound particles can be completely prevented by a large number of nearby bound particles.  This is never true in the hole case, however, because unbound particles can always diffuse through any region of the channel using the interior sites.  Thus, the correlations in the hole case are always negligible, and mean-field theory provides an excellent description of this system.

\section{Discussion and Conclusion}

Diffusive transport driven by concentration gradients can occur under varying degrees of confinement, depending on the relative size of the channel and the diffusing entity. In one extreme, the particle size is much smaller than its surrounding environment, and the transport behavior obeys the standard Fick's law.  In the other extreme, the degree of confinement is severe, forcing the particles to diffuse in a single file.  In this paper, we have presented the first steps toward a systematic understanding of diffusion in systems with an \emph{intermediate degree of confinement}, where steric interactions prevent a large number of particles from diffusing freely past one another.  For symmetric diffusion through a channel with a uniform cross section, we recovered the standard bulk diffusion equation, which neglects steric effects. This is not surprising:  Although steric confinement does affect the single particle dynamics of diffusing particles, it does not alter the bulk diffusive behavior for a symmetric diffusion process, even in the single-file limit~\cite{derrida2}.  We also examined diffusion in a channel with a slowly varying cross section, and were able to derive a generalized form of the Fick-Jacobs equation~\cite{fick, jacobs, zwanzig, reguera, kalinay} for the diffusion of particles in a channel with a varying cross-sectional area.   

In such confined geometries, ubiquitous specific or non-specific interactions between the diffusing objects and the channel walls fundamentally alter the transport dynamics.  When both steric confinement and reversible binding to the channel walls are present, there are two qualitatively different cases that arise as the ratio of the channel width to the particle size is varied. In the first (no-hole) case,  the cross section of the channel is wide enough to accommodate several particles, but still narrow enough that the diffusion of particles through a particular region of the channel can be completely blocked by bound particles in that region. Our simulation results for the bound particle profile indicate a monotonic decrease from the proximal (left) to the distal (right) end of the channel. The steady state current increases as the density of bound particles at the left end increases, but begins to show a saturation behavior at high values, stemming from the fact that the system spends more time in configurations in which the channel is completely blocked. Interestingly, our simulation data for both the particle density profile and the total current at steady state reveal significant deviations from our mean field predictions, especially at higher overall particle densities.  This is due to the fact that binding events that completely block the channel, as well as unbinding events that relieve this blockage, lead to significant deviations in two point density correlation functions from their mean field values. By taking these deviations into account by means of a single dimensionless parameter, we can reproduce both the particle density profiles and the steady state currents seen in all of our simulations. Being able to characterize the particle profile and current across a wide range of concentration gradients across the channel with a single parameter is bound to be extremely useful in predicting transport behavior in systems where a limited amount of data is available.

For the second (hole) case, the channel is wide enough to allow diffusion of particles through its center, even in regions where there is a saturating coverage of bound particles on the wall. Our simulation data in this case is in excellent agreement with the predictions from mean field theory. Since the channel cannot be blocked under any circumstance, the two-point correlations between the bound and unbound particles are much weaker than in the no-hole case, making the deviations of the particle profiles and current from their mean-field values negligible.  Both the particle profiles and the current show qualitative differences from the no-hole case. At high values of the left end particle density, the bound particle profile shows a plateau phase near the left end before dropping to zero at the right end. This indicates that as the concentration gradient across the channel is increased, the particles bind and effectively coat larger and larger regions of the channel wall, starting at the left end.  This is also reflected in the steady state current, which shows a remarkable biphasic behavior as the left end particle density is increased.  At low densities, the current shows only modest increases as the density is raised; at high densities, on the other hand, the current rises sharply for increasing densities. This is due to the aforementioned coating effect of the bound particles at high densities, which forms a non-sticky layer along the channel walls. The diffusive behavior then becomes akin to that of phantom diffusion in a non-sticky channel (albeit of a smaller cross section), resulting in strong increases in the current at high concentrations. This kind of strong biphasic behavior could have important implications in a variety of biological systems, where it could be used as a regulatory or sensory mechanism. In artificial systems, one could potentially tune the system properties to generate a desired strongly non-linear dependence between the current and concentration gradient. Finally, it is important to note that one could go from the no-hole to the hole case with a small change in the channel diameter (on the order of a particle size). That such distinct transport regimes are separated by such small changes in the geometry of the channel could also have wide-ranging implications.
 
Promising avenues for further research include extending our approach to include a bias in the diffusion (i.e. asymmetric diffusion), which could naturally occur as a result of electric fields, hydrodynamic flows or even biased motion of molecular motors. Interactions between the particles themselves could also yield significant new regimes. We hope that our work on these under explored systems, where the intermediate degree of confinement and the particle-wall interactions lead to novel and qualitatively different transport behaviors, inspires further theoretical, computational and experimental research on these very rich systems.

\begin{acknowledgments}
We would like to thank P.~Pincus for stimulating and useful discussions. This work was supported by the NSF under Award No.~DMR-0503347.  The MRL at UCSB is supported by NSF No.~DMR-0080034.  MLH also acknowledges the support of a National Science Foundation Graduate Research Fellowship.  BD acknowledges the support of Lockheed Martin.  Finally, MLH and AG would like to thank the Aspen Center for Physics for providing a stimulating atmosphere in which to complete this work.  
\end{acknowledgments}

\appendix

\section{The Master Equation}
\label{sec:masterEq}

The evolution operator $K$ defined in Eq.~(\ref{eq:masterEq}) can be expressed in terms of the creation and destruction operators $a_b^\pm (i,j)$ and $a_u^\pm (i,j)$ for bound and unbound particles, respectively:
\begin{align}
\nonumber
&a_u^+ (i,j) \left|0; b\right>_{i,j} =  \left|1; b\right>_{i,j} \\
&a_u^- (i,j) \left|1; b\right>_{i,j} =  \left|0; b\right>_{i,j} \\
\nonumber
&a_u^+ (i,j) \left|1; b\right>_{i,j} = a_u^- (i,j) \left|0; b\right>_{i,j}=0,
\end{align}
\begin{align}
\nonumber
&a_b^+ (i,j) \left|u; 0\right>_{i,j} =  \left|u; 1\right>_{i,j}\\
&a_b^- (i,j) \left|u; 1\right>_{i,j} =  \left|u; 0\right>_{i,j} \\
\nonumber
&a_b^+ (i,j) \left|u;1\right>_{i,j} = a_b^- (i,j) \left|u;0\right>_{i,j}=0,
\end{align}
where $\left|u; b\right>_{i,j}$ specifies a state of the site $(i,j)$ with $u$ unbound and $b$ bound particles.  It is important to note that these operators can create the unphysical state $\left|1,1\right>_{i,j}$, in which the site $(i,j)$ is occupied by both a bound and unbound particle.  Therefore, we must take care to construct the evolution operator $K$ so that only transitions between physically allowable states can occur.  This will ensure that, as long as the initial condition of the system is a physically allowable state, the system will never evolve into unphysical states. 

We can construct number operators from the creation and destruction operators, $n_\alpha (i, j) = a_\alpha^+ (i,j) a_\alpha^-(i,j)$, where $\alpha=u,b$.  In terms of these number operators, the operators $C_b(i,j) = n_b(i,j) [1-n_u(i,j)]$ and $C_u(i,j) = n_u(i,j) [1-n_b(i,j)]$.  We also define the operator $C_e(i,j) = [1-n_b(i,j)] [1-n_u(i,j)]$, which gives $1$ if the site $(i,j)$ is empty and $0$ otherwise.

Consider a single event, transforming the system from a state  $\left|s_{old}\right>$ to a state  $\left|s_{new} \right>$,  that is allowed to occur in a time step $\Delta t$ [e.g. the hopping of an unbound particle from $(i,j)$ to $(i+1,j)$]. Using Eqs.~(\ref{eq:masterEq}) and~(\ref{eq:psiDef}) and the orthogonality of the state vectors, $\left<s\right|\left.s'\right>=\delta_{s,s'}$,
\bb
\label{eq:deltaProb}
\Delta \psi [s;t] =  \sum_{\{\tilde{n}_u^{i,j} \}=0,1} \sum_{\{\tilde{n}_b^{i,j} \}=0,1} P[\tilde{s};t] \left<s\right| K \left|\tilde{s}\right>
\en
where for any function $f(t)$, $\Delta f(t) \equiv f(t+\Delta t) -f(t)$.  For every possible transition $\left|s_{old}\right> \rightarrow \left|s_{new}\right>$, there must be two terms in $K$.  Both of these terms should be proportional to $P[s_{old};t]$, since the frequency of the transition $\left|s_{old}\right> \rightarrow \left|s_{new}\right>$ clearly depends on the probability of finding the system in the initial state $\left|s_{old}\right>$. The first term accounts for the \emph{increase} in $P[s_{new};t]$ due to this transition.  This term should give a positive contribution to the RHS of Eq.~(\ref{eq:deltaProb}) for $s=s_{new}$.  Then it is clear from Eq.~(\ref{eq:deltaProb}) that this term should be positive and contain creation and destruction operators that transform $\left|s_{old}\right> \rightarrow \left|s_{new}\right>$. The second term in $K$ for this transition accounts for the \emph{decrease} in $P[s_{old};t]$ due to the transition $\left|s_{old}\right> \rightarrow \left|s_{new}\right>$.  This term should give a negative contribution to the RHS of Eq.~(\ref{eq:deltaProb}) for $s=s_{old}$.  We can see from Eq.~(\ref{eq:deltaProb}) that this term  should be negative and contain \emph{only} number operators, so that the non-zero term in the sum on the RHS of Eq.~(\ref{eq:deltaProb}) is proportional to $P[s_{old};t]$.  Using these rules, it is straightforward to write down the evolution operator $K$ for the system described in section \ref{sec:model}.  Writing $K = \sum_{i,j} K_{i,j}$, 
\begin{widetext}
\begin{align}
\label{eq:Kij}
\nonumber
K_{i,j} =& \Big\{ p_{\rm on} \big[ a_b^+ (i,j) a_u^-(i,j) - n_u(i,j) \left(1-n_b(i,j)\right) \big] + p_{\rm off}\big[ a_b^- (i,j) a_u^+ (i,j) - \left(1-n_u (i,j)\right) n_b (i,j) \big]\Big\} \theta_{j}\\
+& \sum_{j' \neq j}  \tilde{p}_{\rm hop} (j, j') \left(1-n_b(i,j)\right)\left(1-n_b(i,j')\right) \big[ a_u^- (i,j') a_u^+(i,j) - n_u(i,j') \left(1-n_u(i,j)\right) \big]\\
\nonumber
+&  p_{\rm hop} \sum_\pm \left(1-n_b(i,j)\right)\left(1-n_b(i \pm 1,j)\right) \big[ a_u^- (i \pm 1,j) a_u^+(i,j) - n_u(i \pm 1,j) \left(1-n_u(i,j)\right) \big],
\end{align}
\end{widetext}
where $\theta_j=1$ if binding can occur in the $jth$ row, and $0$ if binding cannot occur.  As required, this evolution operator satisfies the constraint that only physically allowable states evolve in time. 

We can now use the master equation Eq.~(\ref{eq:masterEq}) to compute the evolution equations for any given operator.   From Eq.~(\ref{eq:psiDef}), it is clear that the normalization of the state vector $\left|\psi (t)\right>$ is given by $\left<\mathbb{1} \left|\right. \psi (t) \right> = 1$, where $\left| \mathbb{1} \right> \equiv  \sum_s \left|s\right>$.  Therefore, the expectation value of any operator $O_{ij}$ is $\left< O_{ij}\right> \equiv \left<\mathbb{1} \right| O_{ij} \left| \psi (t) \right>$.  Using Eq.~(\ref{eq:masterEq}), the evolution equation for $\left< O_{ij}\right>$ is given by
\bb
\Delta \left< O_{ij} \right> = \left<\mathbb{1} \right| O_{ij} K \left| \psi (t) \right>.
\en
To compute the RHS of this equation, we note that if $K_{i'j'}$ contains no operators that act on the site $(i,j)$, then $\left<\mathbb{1} \right| O_{ij} K_{i'j'} \left| \psi (t) \right> = 0$. Using this fact, it is straightforward to derive the desired evolution equations for the operators of interest.  In particular, the evolution equations for $C_b (i, j)$ and $C_e (i, j)$ are, respectively,  
\bb
\label{eq:CbEvolve}
\Delta \left<C_b (i,j) \right> = \left[p_{\rm on} \left<C_u (i,j) \right>- p_{\rm off} \left<C_b (i,j) \right>\right] \theta_j,
\en
\bb
\label{eq:CeEvolve}
\Delta \left<C_e (i,j) \right> +\Delta \left<C_u (i,j) \right>+\Delta \left<C_b (i,j) \right>= 0.
\en
Since every site must either contain a single particle (bound or unbound), or be unoccupied, and $\left<C_b (i, j)\right>=0$ if $\theta_j =0$, the solution to Eq.~(\ref{eq:CeEvolve}) is simple:
\bb
\left<C_e (i,j) \right>=1-\left<C_u (i,j) \right>- \left<C_b (i,j) \right> \theta_j.
\en
Indeed, for any operator $O_{i'j'}$, 
\bb
\label{eq:Ce}
\left<O_{i'j'}C_e (i,j) \right>=\left<O_{i'j'}\left[1-C_u (i,j)-C_b (i,j) \theta_j\right]\right>.
\en
Using this result, it is straightforward to show that the final desired evolution equation, for the operator $C_u (i,j)$, is given by
\begin{widetext}
\begin{align}
\label{eq:CuEvolve}
\nonumber
\Delta & \left<C_u (i,j) \right> =-\left[p_{\rm on} \left<C_u (i,j) \right>- p_{\rm off} \left<C_b (i,j) \right>\right] \theta_j+ \sum_{j' \neq j} \tilde{p}_{\rm hop} (j,j') \Big[\left<C_u (i,j') \right> - \left<C_u (i,j) \right> - \left<C_u (i,j') C_b (i,j)\right> \theta_j \Big.\\
\Big. & + \left<C_u (i,j) C_b (i,j')\right> \theta_{j'} \Big] + p_{\rm hop} \sum_{\pm} \Big[ \left<C_u (i \pm 1,j) \right> - \left<C_u (i,j) \right> -\left<C_u (i \pm 1,j) C_b (i,j)\right> \theta_j + \left<C_u (i,j) C_b (i \pm 1,j)\right> \theta_j \Big].
 \end{align}
 \end{widetext}


\begin{thebibliography}{99}
 
 \bibitem{ion} B. Hille, \emph{Ion Channels of Excitable Membranes},  (Sinauer, Sunderland, MA, 2001).
 
 \bibitem{chou} T. Chou, Phys. Rev. Lett. {\bf 80}, 85 (1998); T. Chou and D. Lohse, Phys. Rev. Lett. {\bf 82}, 3552 (1999).

\bibitem{porin} H. Nikaido, Microbiol. Mol. Biol. Rev. {\bf 67}, 593, (2003).

 \bibitem{np} D. Gorlich and U. Kutay, Ann. Rev. Cell Dev. Biol. {\bf 15}, 607(1999) ; M. Suntharalingam, S.R. Wente, Dev. Cell. {\bf 4}, 775 (2003). 

\bibitem{odde} D. Odde, {Eur Biophys J} {\bf 27}, 514 (1998).

\bibitem{deborah} J.L. Ross and D. Kuchnir Fygenson, Biophys. J. {\bf 84}, 3959 (2003).  

\bibitem{rbc} A.R. Pries, T.W. Secomb, and P. Gaehtgens, Cardiovasc. Res. {\bf 32}, 654 (1996).

\bibitem{rustom} A. Rustom \emph{et al.}, Science {\bf 303}, 1007 (2004). 

 \bibitem{quake} T.M. Squires,  and S.R. Quake, Rev. Mod. Phys. {\bf 77}, 977 (2005).
 
\bibitem{stone}  H.A. Stone, A.D. Stroock, and A. Ajdari, Ann. Rev. Fluid Mech. {\bf 36}, 381 (2004).

\bibitem{tay} F.E.H. Tay (ed.), \emph{Microfluidics And Biomems Applications} (Kluwer Academic, Netherlands, 2002). 

\bibitem{makamba} H. Makamba \emph{et al.}, Electrophoresis {\bf 24}, 3607 (2003).

\bibitem{hummer} G. Hummer, J. C. Rasaiah, and J. P. Noworyta, Nature {\bf 414}, 188 (2001);  A. Berezhkovskii and G. Hummer, Phys. Rev. Lett. {\bf 89}, 064503 (2002). 

\bibitem{ruthven} J. K\"{a}rger and D. M. Ruthven, \emph{Diffusion in Zeolites and Other Microporous Materials} (Wiley, New York, 1992).

\bibitem{kukla} V. Kukla \emph{et al.}, Science {\bf 272}, 702 (May 3, 1996).

\bibitem{wei} Q.H. Wei, C. Bechinger, P. Leiderer, Science {\bf 287}, 625 (2000).

\bibitem{lin} B.H. Lin \emph{et al.}, Phys. Rev. Lett. {\bf 94}, 216001 (2005).


\bibitem{spitzer} F. Spitzer, Adv. Math. {\bf 5}, 246 (1970).

\bibitem{levitt} D. G. Levitt, Phys. Rev. A {\bf 8}, 3050 (1973).

\bibitem{liggett} T. M. Liggett, \emph{Stochastic Interacting Systems: Contact, Voter, and Exclusion Processes}
(Springer-Verlag, New York, 1999).

\bibitem{derrida} B. Derrida, M.R. Evans, V. Hakim, and V. Pasquier, J. Phys. A: Math. Gen. {\bf 26}, 1493 (1993); B. Derrida, Phys. Rep. {\bf 301}, 65 (1998).

\bibitem{percus}  K.K. Mon and J.K. Percus , J. Chem. Phys. {\bf 117}, 2289 (2002).
 
 \bibitem{kehr} R. Kutner, H. van Beijeren, and K.W Kehr, Phys. Rev. B {\bf 30}, 4382 (1984).
 
\bibitem{kolomeisky} E. Pronina and A.B. Kolomeisky, J. Phys. A: Math. Gen. {\bf 37}, 9907 (2004).

\bibitem{popkov} V. Popkov and I. Peschel, Phys. Rev. E {\bf 64}, 026126 (2001); V. Popkov, J. Phys. A: Math. Gen. {\bf 37}, 1545 (2004).

\bibitem{fick} A. Fick, Poggendorfs Ann. {\bf 94}, 59 (1855).

\bibitem{jacobs} M.H. Jacobs, \emph{Diffusion Processes} (Springer, New York, 1967).

\bibitem{zwanzig} R. Zwanzig, J. Phys. Chem. {\bf 96}, 3926 (1992).

\bibitem{reguera} D. Reguera and J. M. Rub\'{i}, Phys. Rev. E {\bf 64}, 061106 (2001).

\bibitem{kalinay} P. Kalinay and J. K. Percus, Phys. Rev. E {\bf 72} 061203 (2005); \emph{ibid.}, Phys. Rev. E {\bf 74}, 041203 (2006).

\bibitem{derrida2} B. Derrida, J.L. Lebowitz and E.R. Speer, J. Stat. Phys. {\bf 107}, 599 (2002).
 
\bibitem{schultz} G.M. Sch\"utz, J. Stat. Phys. {\bf 88}, 427 (1997).

\bibitem{endnote} This would not be true if $N_H$ was large enough that, for example, some interior rows were surrounded only by other interior rows, with no adjacent exterior rows.


\end{thebibliography}
\end{document}